\documentstyle[12pt]{article}

\begin{document}
\title{ Two-dimensional electron gas
in uniform \\ magnetic and electric fields}
\author{Choon-Lin Ho$^1$, V. R. Khalilov$^2$ and Chi Yang$^1$}
\date{\small $^1$Department of Physics, Tamkang University, Tamsui,
      Taiwan 25137, R.O.C. \\
$^2$ Department of Theoretical Physics, Physical Faculty,\\
Moscow State University, 119899, Moscow, Russia.}

\maketitle

\begin{abstract}

The thermodynamic potential of an ideal nonrelativistic gas of
two-dimensional electrons in crossed uniform
magnetic and electric fields is constructed.
For low temperatures and very weak electric fields, it is shown that
the Hall conductance is always quantized at integral multiples of $e^2/h$
over a large range of strong magnetic fields.
This could be viewed as a quantum statistical explanation of
the integral quantum Hall effect.
Magnetic properties of the system at high temperatures are briefly
discussed.

\end{abstract}


\newpage

Two-dimensional electron gas (2DEG) in external
electromagnetic fields at low
temperature has attracted significant interests
since the discovery of the integral quantum Hall effect (IQHE)
\cite{Prange}.
In IQHE, the Hall conductance
$\sigma_{yx}$ was discovered to have a stepwise dependence (the
appearance of plateaux) on the strong
external magnetic field.  At these plateaux, $\sigma_{yx}$ is quantized
to $\sigma_{yx}=ne^2/h$, $n=1,2,\ldots$ ($e$ is the magnitude of the
electron charge,
and $h=2\pi \hbar$ is the Planck constant) to an accuracy as high as
several parts in $10^8$, while
the longitudinal conductivity $\sigma_{xx}$ nearly vanishes.
The vanishing of $\sigma_{xx}$ implies the absence of dissipation.
The amazing fact that the Hall conductance is quantized to such a high
precision, and that the quantized values are related only to fundamental
constants and are independent of the actual sample and geometry, indicate
that IQHE must be explicable from some general principles independent of
any finer details of the interaction mechanism in the sample.
Efforts in this direction in the past have resulted in theory of IQHE
based on
the principle of gauge invariance \cite{Laughlin} and on topological
considerations \cite{Thouless}.

In view of the fact that IQHE is independent of interaction,
and that it is a  lossless
effect as implied by the vanishing of $\sigma_{xx}$, one may wonder if
IQHE could also be deducible from the quantum statistics of a
non-interacting 2DEG
at thermodynamic equilibrium in the crossed magnetic and electric fields.
The purpose of this paper is to investigate such possibility.

Let us consider a system of nonrelativistic non-interacting 2DEG
in a crossed uniform electric (${\bf E}$) and magnetic
(${\bf B}$) fields, which are specified in Cartesian coordinates as
\begin{eqnarray}
 {\bf E}=(0,\,E,\,0),\ \ {\bf B}=(0,\,0,\,B)=\nabla\times {\bf A},\ \
 {\bf A}=(-yB,\,0,\,0).
\label{e1}
\end{eqnarray}
We shall consider only the case of very weak electric field and strong
magnetic field ($E\ll B$).

The energy spectrum of two-dimensional electron in the field configuration
(1) is given by (see {\sl e.g.} \cite{Landau1,Khalilov}):
\begin{eqnarray}
E_n={\hbar eB\over mc}\left(n+{1\over 2}\right)-c{E\over B}\tilde p+
m_s g {\hbar eB\over 2m_0c}~,\ \ \ m_s=\pm {1\over 2}~,
\label{e2}
\end{eqnarray}
where $n$ is the Landau level number, $m$ is the effective electron
mass in a crystal lattice, $m_0$ is the mass of a free electron,
$g$ is the effective Land\'e
factor, and $\tilde p$ is the momentum of electron in the $x$-direction. For
a very weak electric field, $\tilde p$ is constrained by $|\tilde p|\leq
eBL_y/2c$ \cite{Landau1}.  The last term in (\ref{e2}) describes
the interaction of the spin magnetic moment of the electron with the
magnetic
field.  If the electrons are free we shall put $m=m_0$ and $g=2$.

The main physical quantity which defines all
thermodynamical properties of noninteracting particles gases is the
thermodynamic potential (TP) $\Omega(\tilde\mu,\,B,E,\,T)$, considered
as a function
of the chemical potential ($\tilde\mu$), the external fields
and the gas temperature ($T$).
The TP of the 2DEG in the XOY plane
can be defined as follows (see {\sl e.g.}: \cite{Landau2}):
\begin{eqnarray}
  \Omega (\tilde\mu,\,B,\,E,\,T)=-k_B T
  \sum_{n,\tilde p,m_s}\ln\left[1+\exp\left({\tilde\mu-
  E_n\over k_B T}\right)\right]~,
\label{e3}
\end{eqnarray}
where $k_{\rm B}$ is the Boltzmann constant.
In terms of dimensionless quantities $p\equiv \tilde p/mc$, $\mu\equiv
\tilde\mu/mc^2$, $\varepsilon_n\equiv E_n/mc^2$ and $\theta\equiv
k_B T/mc^2$,
$\Omega$ is given by \begin{eqnarray}
\Omega
(\mu,\,B,\,E,\,\theta)=-mc^2n^{(2)}\theta
\int_{-b/2}^{b/2} dp~{1\over b} \sum_{n,m_s}
  \ln\left[1+\exp\left({\mu-\varepsilon_n\over\theta}\right)\right]~.
\label{e3a}
\end{eqnarray}
Here $L_x,~L_y$ are the lengths of the two-dimensional space,
$b=eBL_y/mc^2$, and
\begin{eqnarray}
  n^{(2)}={eBL_xL_y\over hc}
\label{n2}
\end{eqnarray}
is the number of quantum electron states in the magnetic field for a given
$n$ in an area $L_xL_y$.

Evaluation of the TP is easily carried out by the Mellin transformation
method with respect to the variable $\exp(\mu/\theta)$,
resulting in the following expression
\begin{eqnarray}
  \Omega(\mu,\,\kappa,\,E,\,\theta)=\mp
  mc^2\theta n^{(2)}\sum{\rm Res}{\pi\exp(s\mu/\theta)\over s
  \sin(\pi s)} \phantom{mmmmmmmmmm}  \nonumber   \\
  \times{\cosh[s g^\ast \kappa/2\theta]\over\sinh(s\kappa/2\theta)}
  {\sinh[seEL_y/2mc^2\theta]\over seEL_y/2mc^2\theta}, \phantom{mmmmmmm}
\label{e4} \\   g^\ast= {gm\over 2m_0}~,\ \kappa={B\over B^\ast}~,\
  B^\ast={m^2c^3\over e\hbar}~, \phantom{mmmmmmm}  \nonumber
\end{eqnarray}
where the upper (lower) sign refers to closing the contour to the left
(right) of the imaginary axis for
$\mu>0$ ($\mu<0$).  When $\mu>0$, the sum of residues of (\ref{e4}) should
be calculated
at the poles on both the negative real semiaxis $s=0,\,-1,\,-2,\,-3,\ldots$
(including $s=0$) and the imaginary axis $s_l=2\pi il\theta/\kappa$.
For $\mu<0$, the integration contour
is closed to the right and the sum of residues should be calculated
at the poles $s=1,2,3,\ldots$.

The poles located on the real axis contribute to the ``monotonic part''
of $\Omega$, while the poles $s_l=2\pi il\theta/\kappa$
dominate the behavior of its oscillating part. Since we are
interested in an almost degenerate 2DEG, (\ref{e4}) must be
calculated in the limit $\theta\ll\mu$, in which case $\mu>0$. It is
important
that values of the chemical potentials $\mu$ may exceed zero only for
fermions.  It follows that only thermodynamic quantities
characterizing fermions in a magnetic field at low temperature may
exhibit the oscillating behavior (the so-called Landau oscillations).

Let us first consider the case of low temperature
limit when $\theta\ll\kappa$. Taking the sum at the poles
$s_l=2\pi il\theta/\kappa$, and the pole $s=0$, we obtain TP to the first
order in $\theta/\kappa$ in the form
\begin{eqnarray}
\Omega=\Omega_{\rm mon}+\Omega_{\rm osc}~~,
\phantom{ppppppppppppppppppppppppp}   \nonumber    \\
\Omega_{\rm mon}  =-mc^2n^{(2)}\kappa\left[{z^2\over 4\pi^2}+{1\over 4}
\left(g^{\ast 2}-{1\over 3}\right) - {\gamma \theta\over \pi\kappa}z
 + {\alpha^2\over 12}\right]~,\ \ \label{e5}  \\
\Omega_{\rm osc}  = -mc^2n^{(2)}\kappa \sum_{l=1}^\infty
\left(-1\right)^{l+1}
{\cos(g^\ast\pi l)\cos(z l)\over\pi^2 l^2}
{\sin(\alpha\pi l)\over \alpha\pi l}~,\nonumber
\end{eqnarray}
where  $z=2\pi\mu/\kappa$, $\alpha=eEL_y/\kappa mc^2$ and $\gamma=0.57721$
is
the Euler constant. We note here that in deriving the monotonic part of TP
$(\Omega_{\rm mon})$, we took into account only the contribution of the
pole $s=0$, since the poles $s\ne 0$ contribute to terms of the order
$(-1)^{l+1}\exp\left\{-\left[l(\mu-(g^\ast
-1)\kappa/2\right]/\theta\right\}/l$, which are exponentially small.

In the case of very weak electric fields such that $\alpha\ll 1$,
{\it i.e.} for $E$ satisfying
\begin{eqnarray}
  eEL_y\ll \kappa mc^2~,
\label{E}
\end{eqnarray}
the TP in (\ref{e5}) reduces to
\begin{eqnarray}
\Omega &=& -mc^2n^{(2)}\kappa\left[{z^2\over 4\pi^2}+{1\over 4}
\left(g^{\ast 2}-{1\over 3}\right) - {\gamma \theta\over \pi\kappa}z
 + {1\over \pi^2} F(z)\right]~,\ \ \nonumber \\
F(z) & =&  \sum_{l=1}^\infty
\left(-1\right)^{l+1}{1\over l^2}\cos(g^\ast\pi l)\cos(z l)~.
\label{e6}
\end{eqnarray}

Having derived the TP of the electron gas, we can then calculate the
number of particles
\begin{eqnarray}
  N_e=-{\partial\Omega\over mc^2\partial\mu}~,
\label{e7}
\end{eqnarray}
and the magnetic moment
\begin{eqnarray}
  {\bf M}=-{\partial\Omega\over \partial{\bf B}}
\label{e8}
\end{eqnarray}
of the electron gas in fields (\ref{e1}) at thermodynamic equilibrium.
From (\ref{e6}) and
(\ref{e8}) the magnetic moment of the system is
\begin{eqnarray}
  M_x=M_y=0~\phantom{ppppppppppppppppppppp}
\nonumber     \\
 M_z=mc^2\mu{eL_xL_y\over \hbar c}\left[{1\over 2}\left(g^{\ast 2}-{1\over
3}\right){1\over z} + {2F(z)\over z\pi^2} - {1\over \pi^2}{dF(z)\over dz}
\right]~.
\label{e11}
\end{eqnarray}
One sees that the oscillating part of the magnetic moment dominate in
the limit $\mu\gg\kappa$.  This is the Landau-de Haas-van Alphen effect
\cite{Landau2}.

Now we shall evaluate the number of electrons $N_e$ according to (\ref{e7})
with the TP given by (\ref{e6}).
This quantity is of great importance for the problem under consideration.
In fact, the off-diagonal Hall conductivity in fields (\ref{e1}) can be
related to the number of two-dimensional electrons $N_e$ that occupy quantum
states
with $\varepsilon_n<\varepsilon_{\rm F}$ (where
$\varepsilon_{\rm F}=\tilde\mu$
at $\theta=0$ is the Fermi level) at thermodynamic equilibrium as
follows:
\begin{eqnarray}
  \sigma_{yx}={ecn_e\over B}\equiv{e^2\over h}\cdot {N_e\over n^{(2)}},
\label{e12}
\end{eqnarray}
where $n_e=N_e/L_xL_y$ is the density of electrons.
The ratio
\begin{eqnarray}
\Phi(z) \equiv   {N_e(z)\over n^{(2)}}
\label{Phi}
\end{eqnarray}
as a function of $z$ characterizes the number of occupied Landau
levels, and is usually called the filling factor.
From (\ref{e6}) and (\ref{e7}), we get (for $\theta\ll \kappa$):
\begin{eqnarray}
  \Phi (z)={1\over\pi}\left(z + 2~{dF(z)\over dz}\right)~.
\label{e13}
\end{eqnarray}

The functions $F(z)$ in (\ref{e6}) is a
piecewise periodic function of $z$.  To see this, we express the product
of cosines in (\ref{e6}) as a sum of cosines, and use the identity
\cite{GraRyz}:
\begin{eqnarray}
\sum_{l=1}^\infty (-1)^{l+1} {\cos(lx)\over l^2} ={\pi^2\over 12} -
{x^2\over 4}~,\qquad\qquad\qquad -\pi \leq x \leq \pi~.
\label{e20}
\end{eqnarray}
The result is:
\begin{eqnarray}
F(z)=\cases{{\pi^2\over 12}-{1\over 4}z^2-{1\over 4}g^{\ast 2}\pi^2~,
& $ - (1-g^\ast)\pi\leq  z\leq (1-g^\ast)\pi~$;\cr
{\pi^2\over 12}+{\pi\over  2}z-{1\over 4}z^2~-{1\over 4}g^{\ast 2}\pi^2
-{1\over 2}(1-g^\ast)\pi^2
~,& $(1-g^\ast)\pi\leq z\leq (1+g^\ast)\pi~$.\cr  }
\label{e14}
\end{eqnarray}
Substituting (\ref{e14}) into (\ref{e13}), one finds:
\begin{eqnarray}
\Phi(z)=\cases{ 0~~,& $0\leq z \leq (1-g^\ast)\pi~~$;\cr
                i~~,& $(i-g^\ast)\pi \leq z \leq (i+g^\ast)\pi~~$;\cr
               i+1~~,& $(i+g^\ast)\pi \leq z \leq (i+2-g^\ast)\pi~~$;\cr}
\label{Phi2} \\ \nonumber \\
 i={\rm odd\  integers}=1,3,5,\ldots\phantom{sssssssssssssss}
\nonumber
\end{eqnarray}
Hence the filling factor $\Phi$ assumes only positive integral values.
Consequently, from
(\ref{e12}), we see that the Hall conductance $\sigma_{yx}$ is quantized at
integral multiples of $e^2/h$.
IQHE can thus be explained from the quantum statistical nature of the 2DEG.
The widths of the plateaux corresponding to odd filling factors $\Phi$
are equal to $2 g^\ast\pi$, while those corresponding to even $\Phi$ are
equal to $2 (1-g^\ast) \pi$.
Two special cases, namely, $g^\ast=0$ and
$g^\ast=1$, are worthy of note.  From (\ref{Phi2}) it is clear that when
$g^\ast=0$ the filling factor $\Phi$ is always an even integer, and when
$g^\ast=1$, $\Phi$ is
always odd.  For $g^\ast=0$ there is no splitting of spin degree of
freedom in
each Landau level, and the $\sigma_{yx}$'s are even integral multiples of
$e^2/h$ \cite{Datta}.  For $g^\ast=1$, all Landau levels except the ground
state are spin degenerate, and $\sigma_{yx}$'s are odd integral multiples of
$e^2/h$.  In actual samples exhibiting IQHE, $g^\ast$ are usually
much smaller  than
unity. For instance, $m\approx 0.07 m_0$ and $g\approx 0.8$ in GaAs, giving
$g^\ast\approx 0.03$ \cite{Stomer}.  Thus one expects from (\ref{Phi2}) that
for such small value of $g^\ast$, the
odd plateaux are much narrower than the even ones.  This explains
why the experimentally observed plateaux mostly correspond to even integers
\cite{Cage}.

One could also obtain the quantized values of $\sigma_{yx}$ in layered
multi-quantum-well structures.  These structures contain a number of
well separated identical two-dimensional electronic systems.
If the number of layers in such a structure is $J$, then the total number of
electrons is $N_e^{T}=J N_e$, where $N_e$ is the number of electrons
in a single layer given by $N_e= n^{(2)} \Phi =n^{(2)} I,~I=1,2,3,\ldots$,
according to (\ref{Phi}) and (\ref{Phi2}).  Hence the Hall
conductance of the system is
\begin{eqnarray}
\sigma_{yx}={e^2\over h}\cdot {N_e^{T}\over n^{(2)}}={e^2\over h}IJ~,
\label{multi}
\end{eqnarray}
which agreed with the experimental values \cite{Haa}.

Let us now consider briefly the behaviour of 2DEG at high temperature
($\theta\gg\kappa$)
for a very weak electric field satisfying the condition (\ref{E}), and
for  $E=0$.  In these cases,  the TP is easily
obtained from eq.~(\ref{e4}) by closing the contour to the right
of the imaginary axis and taking the
lower sign.  The residues are evaluated at the poles $s=1,2,\ldots$.  The
result is:
\begin{eqnarray}
  \Omega = mc^2\theta n^{(2)}
\sum_{l=1}^\infty(-1)^l~
  {\exp(l\mu/\theta)\over l}~{\cosh(lg^\ast\kappa/2\theta)\over
\sinh(l\kappa/2\theta)}~.
\label{e16}
\end{eqnarray}

We can also obtain, by the same method described above, the TP of a
two-dimensional nonrelativistic
spin-0 free boson gas at arbitrary temperature.  We give the result here
for reference (see also \cite{May,Toms})
\begin{eqnarray}
  \Omega_0=-{1\over 2}mc^2\theta n^{(2)}
  \sum_{l=1}^\infty
  {\exp(l\mu/\theta)\over l\sinh(l\kappa/2\theta)}~.
\label{e17}
\end{eqnarray}
The factor of $1/2$ accounts for the difference in the spin degree of
freedom. We note here that all the chemical potentials in
(\ref{e16}) and (\ref{e17}) are assumed
to be less than zero. It is worth emphasizing that (\ref{e17}) is valid at
low temperature also, since for bosons one must always close the contour to
the right of the imaginary axis.  From (\ref{e17}), (\ref{e7}) and
(\ref{e8})
we can obtain the boson number and magnetic moment density, and reproduce
the results of \cite{May,Toms} concerning two-dimensional ideal
nonrelativistic spin-0 boson gas at low temperature.

At very high temperature equations (\ref{e16}) and (\ref{e17}) reduce, up
to the first order term in $\kappa /\theta$, to
\begin{eqnarray}
  \Omega   &\cong & mc^2\theta n^{(2)}
  \left[{2\theta\over\kappa}~
  {\rm Li}_2(-e^{\mu/\theta}) -{1\over 4}\left(g^{\ast 2} -{1\over
3}\right)~{\kappa\over \theta}~{1\over \exp(-\mu/\theta)+1}\right]~,
\label{e18} \\
  \Omega_0 & \cong & -{1\over 2}mc^2\theta n^{(2)}
  \left[{2\theta\over\kappa}~
  {\rm Li}_2(e^{\mu/\theta})-{1\over 12}~{\kappa\over \theta}~
  {1\over \exp(-\mu/\theta)-1}\right]~.
\label{e19}
\end{eqnarray}
Here ${\rm Li}_2(x)=\sum_{l=1}^\infty x^l / l^2$ is the dilogarithmic
function of $x$.

We see that the first terms on the right hand sides of (\ref{e18}) and
(\ref{e19}) contribute only in the densities of the particles, while the
second terms give contribution both in the densities of the particles
and the densities of the magnetic moments of the gases.

From (\ref{e18}), (\ref{e19}) and (\ref{e8}) one can check that the
spin-$0$ and spin-$1/2$ (with $g^\ast < 1/\sqrt{3}$) particles
gases produce magnetic moments that direct against the external
magnetic field induction.  They are thus diamagnetic in nature.
When $g^\ast > 1/\sqrt{3}$, the electron gas is paramagnetic.

In summary, we have shown that the two main features of IQHE, namely the
appearance of a sequence of plateaux in the
Hall conductivity $\sigma_{xy}$,
and the exact quntization of
$\sigma_{xy}$ at integral multiples of $e^2/h$, can be
deduced directly in the framework of quantum statistics from an
equilibrium
non-interacting 2DEG in crossed magnetic and electric field in low
temperature.  At high temperature, there exists
a critical value of $g^\ast_{cr}=1/3$ below which the system is
diamagnetic.
Though we have assumed in our derivation a rectangular two-dimensional
space, the results are independent of the geometry of the space in the
thermodynamic limit.

\vskip 1 truecm

\centerline{\bf Acknowlegdment}

This work is supported in part by the R.O.C. Grants number
NSC-85-2112-M-032-002.

\end{document}